Title Page

# Efficient and Visualizable Convolutional Neural Networks for COVID-19 Classification Using Chest CT[1]

Aksh Garg[a] (akshgarg@gmail.com), Sana Salehi[a] (ss_633@usc.edu), Marianna La Rocca[a] (marianna.larocca@loni.usc.edu), Rachael Garner[a] (rachael.garner@loni.usc.edu), Dominique Duncan[a] (dominique.duncan@loni.usc.edu)

[a] Laboratory of Neuro Imaging, USC Stevens Neuroimaging and Informatics Institute, Keck School of Medicine, University of Southern California, 2025 Zonal Avenue, Los Angeles, California, USA

**Corresponding author:** Aksh Garg, Laboratory of Neuro Imaging (LONI), USC Stevens Neuroimaging and Informatics Institute, 2025 Zonal Avenue, Los Angeles, California, USA

E-mail: akshgarg@gmail.com

Tel: (+1) 424-330-5980

The authors declare no conflicts of interest.

---

[1] All figures should appear in color online only.

**ABSTRACT**
With COVID-19 cases rising rapidly, deep learning has emerged as a promising diagnosis technique. However, identifying the most accurate models to characterize COVID-19 patients is challenging because comparing results obtained with different types of data and acquisition processes is non-trivial. In this paper we designed, evaluated, and compared the performance of 20 convolutional neutral networks in classifying patients as COVID-19 positive, healthy, or suffering from other pulmonary lung infections based on Chest CT scans, serving as the first to consider the EfficientNet family for COVID-19 diagnosis and employ intermediate activation maps for visualizing model performance. All models are trained and evaluated in Python using 4173 Chest CT images from the dataset entitled "A COVID multiclass dataset of CT scans," with 2168, 758, and 1247 images of patients that are COVID-19 positive, healthy, or suffering from other pulmonary infections, respectively. EfficientNet-B5 was identified as the best model with an F1 score of $0.9769\pm0.0046$, accuracy of $0.9759\pm0.0048$, sensitivity of $0.9788\pm0.0055$, specificity of $0.9730\pm0.0057$, and precision of $0.9751\pm 0.0051$. On an alternate 2-class dataset, EfficientNetB5 obtained an accuracy of $0.9845\pm0.0109$, F1 score of $0.9599\pm0.0251$, sensitivity of $0.9682\pm0.0099$, specificity of $0.9883\pm0.0150$, and precision of $0.9526 \pm 0.0523$. Intermediate activation maps and Gradient-weighted Class Activation Mappings offered human-interpretable evidence of the model's perception of ground-class opacities and consolidations, hinting towards a promising use-case of artificial intelligence-assisted radiology tools. With a prediction speed of under 0.1 seconds on GPUs and 0.5 seconds on CPUs, our proposed model offers a rapid, scalable, and accurate diagnostic for COVID-19.

# 1. INTRODUCTION

The reverse transcription-polymerase chain reaction, RT-PCR, is currently considered to be the gold-standard for COVID-19 diagnosis. However, the rapid increase in COVID-19 cases, delay in obtaining PCR results, and strict requirements for testing environments make the fast and effective screening of suspected cases challenging (Islam et al., 2021) Moreover, PCR's low sensitivity, or high false-negative rate, results in many COVID-19 positive patients incorrectly being diagnosed as negative, further exacerbating the disease spread. In particular, a recent study by Feng et al. revealed a specificity of 71% for PCR tests, much lower than targeted specificity for effectively containing the spread of the virus (Fang et al., 2020).

Radiological imaging via X-ray radiography and computed tomography (CT) has emerged as a promising alternative form of diagnosis due to its ability to visualize lung structures. Imaging already serves as a quintessential factor by triaging confirmed COVID-19 cases on the basis of the severity of lung involvement (Dong et al., 2020). However, manual readings of scans are prone to error and time-consuming. The use of machine learning (ML) and artificial intelligence (AI) algorithms that can learn from data without the need for explicit programming offers a promising avenue for meeting the high costs and radiologist shortages surrounding CT imaging. While human readings of CT scans can take upwards of 15 minutes, ML-based algorithms can analyze images within a few seconds. Moreover, with developments in computer vision and computational resources, state-of-the-art convolutional neural network (CNN) architectures may reach specificities, sensitivities, and accuracies of as high as 0.992, 1.00, and 0.995, respectively in distinguishing between COVID-19 and Non-COVID-19 lung CT images (Islam et al., 2021).

With the diverse array of available models for diagnosis, identifying the most optimal has become a valued yet incredibly challenging task. Although several literature reviews consider the use of ML and AI for COVID-19 diagnosis and severity assessment, they presented models trained on different datasets, evaluated with varying metrics, and focusing on different goals – binary classification vs. multiclass classification, classification vs. segmentation, etc (Islam et al., 2021; Lalmuanawma et al., 2020; Ozturk et al., 2020; Waleed Salehi et al., 2020; Xiong et al., 2020). In contrast, our paper presents 20 ML models trained on a fixed dataset, evaluates their performance through metrics such as specificity, sensitivity, accuracy, F-1 scores, and applies visualization techniques such as Gradient-weighted Class Activation Mappings (GradCAMs) and intermediate activation maps to highlight core features such as ground-glass opacities, consolidations, crazy paving patterns, and linear opacities in the input CT images that the model used for making predictions.

Moreover, EfficientNets, with their markedly smaller network sizes and extremely high accuracies in the ImageNet dataset, (Deng et al., 2010) have rapidly become a go-to choice for image-recognition tasks with ML. However, to the best of our knowledge, this paper is the first to consider the entire EfficientNet family of CNN architectures for diagnosis on CT images. Although a limited number of studies have directed their attention to this nascent CNN architecture, they restrict their consideration to chest X-ray images (Chowdhury et al., 2021; Marques et al., 2020; Muftuoglu et al., 2020). While X-ray radiography is cheaper and more universally accessible, CT imaging is preferred over X-ray for diagnoses because of its detailed cross-sectional images (Kim et al., 2020; Ye et al., 2020). Moreover, the yet fewer studies which train EfficientNets on CT Scan images limit their study to EfficientNet B3 and EfficientNet B4, leaving the remaining models EfficientNet B0, EfficientNet B1, EfficientNet B2, EfficientNet B5, EfficientNet B6, and EfficientNet B7 unexplored in terms of their COVID-19 diagnosing abilities (Xiong et al., 2020; Yousefzadeh et al., 2020). Given that many of the larger

EfficientNet architectures acquire the highest accuracy on the ImageNet dataset, this study includes them for comparative purposes and hopes of attaining higher performance.

Finally, this paper is the first to visualize intermediate activation maps for COVID-19 diagnosis. Although the conventional visualization framework —GradCAMs— are useful for localizing abnormalities in input images, they do not offer insight into the model's learning process. In contrast, intermediate activations help understand how successive CNN filters transform their inputs and get a more thorough understanding of individual CNN filters and the model learning behavior (Francois Chollet, 2017).

In summary, the contributions of our paper are multifaceted. Firstly, we offer a standardized basis for comparing 20 state-of-the-art neural network architectures, a feat infeasible to accomplish given the variation in datasets used, goals, and metrics reported. Secondly, we popularize the use of EfficientNets by including the entire family for classification purposes and demonstrating the improvements it offers for COVID-19 diagnosis. Finally, we propose a new visualization mechanism involving both Intermediate Activation Maps and GradCAMs. These allow users to both observe key infection regions within the lung that the ML models used for diagnosis and dispel the stigma surrounding the black-box nature of ML algorithms by offering insight into the model learning process.

The rest of this paper is organized as follows: Section 2 delves deeper into recent works regarding using deep learning techniques for COVID-19, highlighting the volume of research within the field and remaining gaps of weakness. Section 3 introduces the methodology used for dataset accumulation and processing, model training and evaluation, and visualization. Section 4 presents the testing results from the models trained. Section 5 presents the visualizations from GradCAMs. Section 6 presents the visualization results from the Intermediate Activation Maps.

Section 7 presents the neural network architecture for the best performing model. Section 8 discusses and analyzes the findings as well as acknowledging any limitations. Finally, section 9 concludes the paper, summarizing its findings, and suggesting directions for future work.

## 2. RELATED WORKS

Since the introduction of deep learning-based techniques for COVID-19 in a work by Wu et al., several works have been dedicated to evaluating their efficacy (X. Wu et al., 2020). For example, Butt et al. considered the use of ResNet18 attaining an accuracy of 0.867, sensitivity of 0.815, precision of 0.808, and F1 score of 81.1 (Butt et al., 2020). In (Wang et al., 2021) and (Jin et al., 2020), the authors trained and evaluated ResNet152, DPN-92, Inception-v3, ResNet50, and Attention ResNet-50 with U-Net++, reaching accuracies and sensitivities as high as 94.98 and 94.06, respectively. Similarly works by Yousefzadeh et al. and Aradakani et al. extended these efforts further by collectively training DenseNets, Xception, EfficientNetB0, AlexNet, VGG-16, VGG-19, Squeeze Net, Google Net, and MobileNet-V2 for COVID-19 diagnosis, reaching sensitivities as high as 1.00 and accuracies as high as 0.9951 for diagnosis (Ardakani et al., 2020; Yousefzadeh et al., 2020).

More recently, works have instead focused on developing novel machine learning pipelines for COVID-19 classification. For instance, (Foysal & Aowlad Hossain, 2021) developed an ensemble of shallow CNNs to distinguish between COVID-19 positive and negative images, attaining accuracies and sensitivities of .96 and 0.97 respectively. (Ibrahim et al., 2021) developed a modified version of VGG16 – Norm-VGG16 – which attains an accuracy and sensitivity of 0.978 and 0.967, respectively. (Oyelade et al., 2021) propose a new deep learning framework – CovFrameNet – that attains a recall of 0.85, F1 score of 0.9, and specificity of 1.0 in detecting COVID-19. Alrahlal and KP developed a fusion of ResNet-50 and gradient-boosting methods to classify COVID-19/Healthy with an accuracy of 0.9784 (Alrahhal & K P, 2021). Singh and Kolekar attempted to address the computational expensiveness of deep learning with a low-latency MobileNet model with an accuracy of 0.964 (Singh & Kolekar, 2021). Chaudhary and Pachori introduced a Fourier-Bessel series decomposition method, which when combined when ResNet50 attained accuracies of 0.976 and sensitivity of 0.97 (Chaudhary & Pachori, 2021). Table 1 summarizes the performance of recent works involving deep learning for COVID-19 classification. Garg et al. tackle a 3-class classification problem using ResNet50, attaining a testing accuracy of 88.89 (Garg et al., 2021). Li et al. use stacked generalization ensemble learning with VGG16, attaining an accuracy, sensitivity, specificity, precision, F1 score of 0.9357, 0.9421, 0.9393, 0.8940, and 0.9174, respectively (Li et al., 2021). Finally, Garain et. al used a spiking neural network based approach for classification, attaining a F1 score of 0.72 and precision of 0.63 (Garain et al., 2021).

Table 1 provides a summary of recent deep learning methods for COVID-19 diagnosis. Overall, the table and this section highlight two core features: (1) the large variation and lack of consensus regarding which base architecture to use for COVID-19 diagnosis and (2) the dearth of works that use EfficientNet for diagnosis. By removing confounding factors stemming from dataset variations and fully examining the EfficientNet class, our works establishes a consistent backbone for future works.

**TABLE 1: SUMMARY OF RECENT DEEP LEARNING EFFORTS FOR COVID-19 CLASSIFICATION FROM CHEST IMAGES**

| AUTHORS | Mode | Methods | Classes | Metrics |
|---|---|---|---|---|
| (ALRAHHAL & K P, 2021) | CT | ResNet50 + AdaBoost | 2 | Accuracy: 0.9784 |
| (ARDAKANI ET AL., 2020) | CT | AlexNet, VGG-16, VGG-19, SqueezeNet, GoogleNet, MobileNet-V2, ResNet-18, ResNet-50, ResNet-101, Xception | 2 | Accuracy: 0.9951 Sensitivity: 1.00 Specificity: .9902 |
| (BOUGOURZI ET AL., 2021) | CT | ResNet-50, DenseNet161, Inception-V3, Wide-ResNet + XGBoost | 3 | Overall Accuracy: 0.8775 Covid Sensitivity: 0.9636 Pneumonia sensitivity: 0.5263 Normal Sensitivity: 0.9583 |
| (CHAUDHARY & PACHORI, 2021) | Xray | Fourier-Bessel Series Decomposition + ResNet-50, AlexNet, NASNet, EfficientNet, Inception ResNet-v2 | 2 | Accuracy: 0.976, Sensitivity: 0.97 |
| (CHOWDHURY ET AL., 2021) | X-ray | EfficientNet-B1-5 | 2 | Overall Accuracy: 0.9607 Recall (COVID Positive): 1.00 |
| (FOYSAL & AOWLAD HOSSAIN, 2021) | CT | Ensemble of 3 Deep CNNs | 2 | Accuracy: 0.96 Sensitivity: 0.97 |
| (GARAIN ET AL., 2021) | CT | Spiking Neural Network | | F1 Score: 0.74, Precision: 0.63 Recall: 0.92 |
| (GARG ET AL., 2021) | CT | Multi-Scale Residual Network + Ensemble Classifier | 3 | Accuracy: 0.8889 |
| (IBRAHIM ET AL., 2021) | CT | Norm VGG16 + Hand Crafted Features | 2 | Accuracy: 0.978 Sensitivity: 0.967 |
| (JIN ET AL., 2020) | CT | ResNet152 + UNet ++ | 3 | AUC: 0.9299 |
| (KAMEL ET AL., 2021) | CT | Global Thresholding + VGG19 | 2 | Accuracy: 9831, Recall: 1.00, Precision: .9819, F1-score: 0.9864 |
| (KAYA ET AL., 2021) | CT | VGG-16, EfficientNetB3, ResNet50, MobileNetv2 | 2 | Accuracy: 0.979 |
| (LI ET AL., 2021) | CT | Stacked Generalization Learning + VGG16 | 3 | Accuracy: 0.9357, Sensitivity: 0.942 Specificity: 0.9393 |
| (MARQUES ET AL., 2020) | X-ray | EfficientNetB4 | 3 | Overall Accuracy: 0.9670 Recall: 0.9669 |
| (MUFTUOGLU ET AL., 2020) | X-ray | Differential Privacy Practice via EfficientNet-B0 | 2 | Accuracy: 0.947 |
| (OYELADE ET AL., 2021) | CT | CovFrameNet: Pipeline Image Preprocessing + Deep Neural Network Classification | 2 | Recall: 0.85; F1-Score: 0.90 Specificity: 1.0 |
| (OZTURK ET AL., 2020) | Xray | DarkCovidNet | Both | Multi-Class Accuracy: 0.8702 Binary Accuracy: 0.9808 |
| (SINGH & KOLEKAR, 2021) | CT | MobileNetv2 | 2 | Accuracy: 0.9640 |
| (WANG ET AL., 2021) | CT | ResNet152, DPN-92, Inception-v3 | 2 | Accuracy: 0.9498 Sensitivity: 0.9406 |
| (X. WU ET AL., 2020) | CT | ResNet50 | 2 | AUC: 0.819, Accuracy: 0.760 Sensitivity: 0.811, Specificity: 0.615 |
| (Y. H. WU ET AL., 2021) | CT | Joint Classification and Segmentation | 2 | Sensitivity: 0.950, Specificity: 0.930 |
| (XIONG ET AL., 2020) | CT | EfficientNetB4 | 2 | Accuracy: 0.87; Sensitivity: 0.89; Specificity: 0.86; |
| (YOUSEFZADEH ET AL., 2020) | CT | EfficientNet-B3 | 2 | AUC: 0.954, |

## 3. METHODS

### 3.1. Dataset

A dataset containing 4173 CT images of 210 different patients was obtained from a dataset entitled "A COVID multiclass dataset of CT scans" on Kaggle (Soares, Eduardo (Universidad de Lancaster); Angelov, 2020). The dataset may further be triaged into 3 categories, comprised of 2168 images of 80 patients infected with COVID-19 (~27 images/person), 758 images of 50 healthy patients (~15 images/person), and 1247 images of 80 patients with other pulmonary infections (~20 images/person). All images were grayscale in nature, collected from patients in Sao Paulo, Brazil, and made freely accessible through Kaggle by Soares E. et al (Soares, Eduardo (Universidad de Lancaster); Angelov, 2020).

### 3.2. Data Preprocessing

Data preprocessing is an essential step in ML because a model learns to recognize patterns based on the data that it receives. To prevent data leakage from the training dataset to test dataset, we adopted a patient-wise split rather than an image-wise split. Particularly, we first separated all patients into preliminary training and testing sets via 5-fold cross-validation. Subsequently, all images belonging to the patients in the preliminary training set were assigned to the training set and all images corresponding to patients in the preliminary testing set were assigned to the test set. This ensured that no images belonging to the same patient were present in both the training and test set, thereby removing any model confounding that may occur from a CNN learning a patient's chest shape or lung structure. Thus, this step ensured CNN classifications were attributable to pathologies within the lung alone. Secondly, additional preprocessing steps were introduced to scale the generalizability of the proposed models (an analysis of performance on alternate datasets is presented in section V). Particularly, all images in the training dataset were augmented during run-time through the addition of random rotations, horizontal shifts, vertical shifts, skews, and sheers through built-in functions in Keras Image Data Generators. Finally, all images in the training dataset were shuffled to increase variance as the model advanced from one image to the next. A quantitative summary of the augmentations applied is presented in Table 2.

Table 2: Summary of Data Augmentation Techniques Applied

| Transformation | Range |
| --- | --- |
| Pixel Rescaling Factor | 1/255 |
| Horizontal Flips Allowed | True |
| Vertical Flips Allowed | True |
| Zoom Range | [0.85, 1.15] |
| Rotation Range | [0º, 360º] |
| Width Shift Range | [-15%, 15%] |
| Height Shift Range | [-15%, 15%] |
| Shear Range | [-15%, 15%] |

### 3.3. Model Development and Training

A total of 20 models were trained and evaluated for the purposes of this study. These models were derived from the following base models: EfficientNet B0 (Tan & Le, 2019), EfficientNet B1 (Tan & Le, 2019), EfficientNet B2 (Tan & Le, 2019), EfficientNet B3 (Tan & Le, 2019), EfficientNet B4 (Tan & Le, 2019), EfficientNet B5 (Tan & Le, 2019), EfficientNet B6 (Tan & Le, 2019), EfficientNet B7 (Tan & Le, 2019), ResNet 50 (He et al., 2016b), ResNet 50V2 (He et al., 2016a), ResNet 101V2 (He et al., 2016b), ResNet 152V2 (He et al., 2016b), InceptionV3 (Szegedy et al., 2016), InceptionResNetV2 (Szegedy et al., 2017), Xception (François Chollet, 2017), DenseNet 121 (Huang et al., 2016), DenseNet 169 (Huang et al., 2016), DenseNet 201 (Huang et al., 2016), VGG16 (Karen Simonyan∗ & Andrew Zisserman+, 2018), and VGG19 (Karen Simonyan∗ & Andrew Zisserman+, 2018). A summary of each model family is presented in the Appendix. Many of these models have obtaining state-of-the-art performance on the benchmark ImageNet dataset (Deng et al., 2010). Given their exemplary performance, the general model architecture was retained, and the weights initialized using their versions from ImageNet. However, the final output SoftMax layer was changed from a 1000-dimensional to 2-dimesnsional node to make the models compatible for classification between COVID-19 positive images, healthy images, and images from other pulmonary infections. Even though the network architecture was kept the same, the entire model weights were trained from scratch using a Tesla V100-SXM2-32GB GPU on TensorFLow 2.3.0. This allowed us to ensure each memory unit of the proposed model architecture was fine-tuned and beneficial in classification, rather than a wasteful transmission of information that may often occur in transfer learning – especially in a situation like ours where images from ImageNet might not adapt well to CT images.

A pilot study on a subset of the original dataset's images was conducted to identify the optimal hyperparameters to use for model training. First, a custom loss function using label smoothing on top of categorical cross entropy was used to train the network. By transforming the otherwise hard class label assignments (0: COVID, 1: Healthy, 2: Others) into soft label assignments, it reduced model overfitting and increased its likelihood of generalizing better. To perform gradient descent on our model, we relied upon the Adam Optimizer with a learning rate of 0.0001. Next, we introduced a reduce learning rate on plateau callback, which decreased the optimizer learning rate by a factor of 0.5 after 3 consecutive epochs where the increases in performance were 0.0001 or less. This increased our model's capacity to converge to the true local minimum as learning stagnated. Finally, we monitored the performance of the model during each epoch by using a validation split of [85,15] in the training data. A reference to the best performing model state across all epochs was maintained and used for calculating the performance on the testing dataset.

The performance of additional network modifications, including model layers, dropout, and batch normalization, was experimentally tested in a series of pilot studies. However, modifications typically resulted in performance reductions and consequently, we limited our focus on core model families.

### 3.4. Model Evaluation

Each model was trained and validated by running 25 rounds of 5-fold cross-validation. The accuracy, specificity, sensitivity, precision, and F1 scores for each class were subsequently found. The average value of the metrics over all rounds was then computed and their expected values presented within a 95% confidence interval. A description of the metrics is below:

$$Accuracy = \frac{TP + TN}{TP + TN + FP + FN} \tag{1}$$

$$Precision = \frac{TP}{TP + FP} \tag{2}$$

$$Specificity = \frac{TN}{TN + FP} \tag{3}$$

$$Sensitivity\ or\ Recall = \frac{TP}{TP + FN} \tag{4}$$

$$F1\ Score = 2 * \frac{Precision * Recall}{Precision + Recall} \tag{5}$$

Where,

1) True Positive (TP) represents the model correctly classifying an image from a particular class as that class.

2) True Negative (TN) represents the model correctly classifying an image not belonging to a particular class as not being from that class.

3) False Positive (FP) represents the model incorrectly classifying an image not belonging to a particular class as belonging to that class.

4) False Negative (FN) represents the case when a model incorrectly classifies a model belonging to a particular class as not belonging to that class.

Several factors were considered when identifying the more appropriate metrics by which to rank the model performance. In the case of COVID-19 diagnosis, failing to classify a COVID-19 patient as having the disease allows the disease to spread rapidly, exposing a greater number of patients at risk. In contrast, if a COVID-19 negative patient is classified as positive, the error, albeit time and cost-invasive, may easily be corrected in subsequent testing through PCR. Therefore, while all results are presented, this paper attributes the greatest emphasis on the sensitivity, i.e., the model's ability to diagnose a COVID-19 positive patient as having the disease correctly.

### 3.5. Visualization

Intermediate Activation Maps and GradCAMs (Selvaraju et al., 2020) were used to identify which portions of the images the model is using to make diagnoses. These visualizations become especially important when considering the general stigma against ML and CNN's black-box nature. By offering human-interpretable insight into the procedures, the model performs while making diagnoses, they effectively increase the chances of being received favorably by human evaluators.

### 3.5.1. GradCAMs

The GradCAMs were computed using the process outlined in (Selvaraju et al., 2020). First, we found the neuron importance weights:

$$\omega_k^c = \frac{1}{Z} \sum_i \sum_j \frac{\partial Y^c}{\partial A_{i,j}^k} \quad (6)$$

Where $A_{i,j}^k$ represents the activation map of the $k$th filter of a convolutional layer and $Y^c$ represents the probability of classifying class c. These weights were then combined with the forward activation maps in a weighted manner and then passed through a ReLU filter to obtain the class discriminative saliency maps for the targeted image c.

$$L_{i,j}^c = ReLU * \sum_k \omega_k^c A_{i,j}^k \qquad (7)$$

After generating the GradCAMs, we computed the mean intensities of the RGB-pixels of the generated heatmaps corresponding to regions associated with high neuron importance weights that we found by calculating the mean pixel intensity value for areas in the entire heatmap image with intensities one standard deviation above the mean as part of the mask.

The masks were first found using the base model's heatmap image as a template. The generated masks were then applied to both the base model and the modified-network to compare localization and visualization abilities between the base models and modified-networks.

### 3.5.2. Feature Map and Filter Visualization (Intermediate Activation Maps)

To visualize intermediate feature maps for the network, we simply iterated through layers in between the input and output layer of the CNN and extracted the pixel values from each filter's outputs. Over time the model learned more specific features from the images, moving from output representations like the input image to gradually towards drastically different ones.

Intermediate activation maps operated from the principle that each filter in a CNN learns different features. For example, the first filter in the opening layer may be detecting vertical edges, horizontal edges, or color gradients. As one moves deeper into the network, the results from applying the filter from the preliminary layer generated a new array of pixels representing a new "intermediate" image. This is passed to the filter of the next layer, and over time the model can learn more complex embeddings. However, the filters themselves were not manually designed but instead learned by the model through the training process. Thus, they offered a tremendous amount of insight into how the steps undertaken by a model from going from an input image to its final classification.

## 4. RESULTS

### 4.1. COVID-19 Positive Classification

Table 3 summarizes the performance for all trained models in classifying COVID-19 positive images. EfficientNetB5 attained the highest F1 score, accuracy, and sensitivity, while DenseNet169 obtained the greatest specificity and precision.

*Table 3: Summary of model performance for COVID-19 classification. The best performing model in each metric is highlighted in green. EfficientNetB5 attained the greatest F1 score, accuracy, and sensitivity, whereas DenseNet169 obtained the highest specificity and precision.*

| # | Model | F1 | Accuracy | Sensitivity | Specificity | Precision |
|---|---|---|---|---|---|---|
| 1 | DenseNet121 | 0.9709 ± 0.0059 | 0.9699 ± 0.0061 | 0.9655 ± 0.0103 | 0.9747 ± 0.0054 | 0.9767 ± 0.0047 |
| 2 | DenseNet169 | 0.9729 ± 0.0065 | 0.9719 ± 0.0067 | 0.9683 ± 0.0101 | 0.9759 ± 0.0072 | 0.9779 ± 0.0063 |
| 3 | DenseNet201 | 0.9733 ± 0.0058 | 0.9723 ± 0.0061 | 0.9703 ± 0.0090 | 0.9743 ± 0.0063 | 0.9766 ± 0.0054 |
| 4 | EfficientNetB0 | 0.9648 ± 0.0051 | 0.9633 ± 0.0054 | 0.9658 ± 0.0083 | 0.9608 ± 0.0080 | 0.9644 ± 0.0069 |
| 5 | EfficientNetB1 | 0.9300 ± 0.0250 | 0.9276 ± 0.0244 | 0.9323 ± 0.0298 | 0.9226 ± 0.0361 | 0.9350 ± 0.0260 |
| 6 | EfficientNetB2 | 0.9546 ± 0.0062 | 0.9530 ± 0.0066 | 0.9476 ± 0.0090 | 0.9590 ± 0.0080 | 0.9622 ± 0.0070 |
| 7 | EfficientNetB3 | 0.9594 ± 0.0065 | 0.9580 ± 0.0066 | 0.9552 ± 0.0106 | 0.9613 ± 0.0070 | 0.9642 ± 0.0064 |
| 8 | EfficientNetB4 | 0.9647 ± 0.0072 | 0.9634 ± 0.0074 | 0.9637 ± 0.0113 | 0.9635 ± 0.0074 | 0.9663 ± 0.0068 |
| 9 | EfficientNetB5 | 0.9769 ± 0.0046 | 0.9759 ± 0.0048 | 0.9788 ± 0.0055 | 0.9730 ± 0.0057 | 0.9751 ± 0.0051 |
| 10 | EfficientNetB6 | 0.9614 ± 0.0053 | 0.9597 ± 0.0056 | 0.9661 ± 0.0080 | 0.9532 ± 0.0088 | 0.9573 ± 0.0078 |
| 11 | EfficientNetB7 | 0.9448 ± 0.0074 | 0.9432 ± 0.0077 | 0.9397 ± 0.0131 | 0.9475 ± 0.0087 | 0.9511 ± 0.0077 |
| 12 | InceptionResNetV2 | 0.9450 ± 0.0069 | 0.9427 ± 0.0074 | 0.9464 ± 0.0124 | 0.9392 ± 0.0097 | 0.9443 ± 0.0083 |
| 13 | InceptionV3 | 0.9567 ± 0.0070 | 0.9549 ± 0.0072 | 0.9587 ± 0.0117 | 0.9509 ± 0.0099 | 0.9554 ± 0.0087 |
| 14 | ResNet101V2 | 0.9383 ± 0.0107 | 0.9364 ± 0.0116 | 0.9289 ± 0.0156 | 0.9450 ± 0.0151 | 0.9490 ± 0.0128 |
| 15 | ResNet152V2 | 0.9407 ± 0.0099 | 0.9380 ± 0.0107 | 0.9441 ± 0.0158 | 0.9315 ± 0.0185 | 0.9389 ± 0.0139 |
| 16 | ResNet50 | 0.9638 ± 0.0061 | 0.9625 ± 0.0062 | 0.9609 ± 0.0104 | 0.9643 ± 0.0093 | 0.9672 ± 0.0084 |
| 17 | ResNet50V2 | 0.9335 ± 0.0092 | 0.9308 ± 0.0099 | 0.9328 ± 0.0154 | 0.9292 ± 0.0185 | 0.9361 ± 0.0143 |
| 18 | VGG16 | 0.8932 ± 0.0107 | 0.8889 ± 0.0111 | 0.8954 ± 0.0190 | 0.8828 ± 0.0166 | 0.8930 ± 0.0136 |
| 19 | VGG19 | 0.8673 ± 0.0189 | 0.8558 ± 0.0304 | 0.8838 ± 0.0219 | 0.8247 ± 0.0748 | 0.8599 ± 0.0337 |
| 20 | Xception | 0.9491 ± 0.0062 | 0.9470 ± 0.0064 | 0.9510 ± 0.0118 | 0.9432 ± 0.0112 | 0.9482 ± 0.0096 |

## 4.2. Healthy Image Classification

Table 4 summarizes the performance for all trained models in classifying healthy images. EfficientNetB5 attained the highest F1 score, accuracy, sensitivity, specificity, and precision.

Table 4: Summary of model performance for healthy image classification. The best performing model in each metric is highlighted in green. EfficientNetB5 attained the highest F1 score, accuracy, sensitivity, specificity, and precision.

|  | Model | F1 | Accuracy | Sensitivity | Specificity | Precision |
|---|---|---|---|---|---|---|
| 1 | DenseNet121 | 0.7835 ± 0.0263 | 0.9193 ± 0.0099 | 0.8043 ± 0.0334 | 0.9445 ± 0.0102 | 0.7700 ± 0.0350 |
| 2 | DenseNet169 | 0.7835 ± 0.0277 | 0.9195 ± 0.0102 | 0.8045 ± 0.0365 | 0.9448 ± 0.0100 | 0.7704 ± 0.0330 |
| 3 | DenseNet201 | 0.7854 ± 0.0281 | 0.9197 ± 0.0110 | 0.8055 ± 0.0302 | 0.9449 ± 0.0111 | 0.7723 ± 0.0385 |
| 4 | EfficientNetB0 | 0.7909 ± 0.0247 | 0.9201 ± 0.0099 | 0.8259 ± 0.0271 | 0.9410 ± 0.0095 | 0.7651 ± 0.0321 |
| 5 | EfficientNetB1 | 0.7307 ± 0.0492 | 0.8965 ± 0.0176 | 0.7875 ± 0.0554 | 0.9207 ± 0.0183 | 0.7041 ± 0.0445 |
| 6 | EfficientNetB2 | 0.7949 ± 0.0241 | 0.9197 ± 0.0099 | 0.8481 ± 0.0274 | 0.9357 ± 0.0102 | 0.7551 ± 0.0315 |
| 7 | EfficientNetB3 | 0.7912 ± 0.0233 | 0.9194 ± 0.0095 | 0.8332 ± 0.0268 | 0.9387 ± 0.0098 | 0.7606 ± 0.0321 |
| 8 | EfficientNetB4 | 0.7925 ± 0.0276 | 0.9220 ± 0.0106 | 0.8193 ± 0.0335 | 0.9448 ± 0.0097 | 0.7744 ± 0.0322 |
| 9 | EfficientNetB5 | **0.8217 ± 0.0249** | **0.9322 ± 0.0109** | **0.8488 ± 0.0185** | **0.9504 ± 0.0118** | **0.8010 ± 0.0383** |
| 10 | EfficientNetB6 | 0.7891 ± 0.0242 | 0.9177 ± 0.0099 | 0.8465 ± 0.0291 | 0.9337 ± 0.0093 | 0.7448 ± 0.0301 |
| 11 | EfficientNetB7 | 0.7810 ± 0.0243 | 0.9161 ± 0.0100 | 0.8169 ± 0.0265 | 0.9383 ± 0.0096 | 0.7538 ± 0.0315 |
| 12 | InceptionResNetV2 | 0.7727 ± 0.0267 | 0.9154 ± 0.0102 | 0.7918 ± 0.0370 | 0.9429 ± 0.0104 | 0.7622 ± 0.0360 |
| 13 | InceptionV3 | 0.7673 ± 0.0295 | 0.9130 ± 0.0112 | 0.7912 ± 0.0405 | 0.9402 ± 0.0115 | 0.7541 ± 0.0382 |
| 14 | ResNet101V2 | 0.7408 ± 0.0347 | 0.9035 ± 0.0116 | 0.7677 ± 0.0448 | 0.9333 ± 0.0108 | 0.7234 ± 0.0364 |
| 15 | ResNet152V2 | 0.7596 ± 0.0329 | 0.9093 ± 0.0120 | 0.7931 ± 0.0423 | 0.9351 ± 0.0115 | 0.7367 ± 0.0384 |
| 16 | ResNet50 | 0.7784 ± 0.0253 | 0.9174 ± 0.0101 | 0.7961 ± 0.0290 | 0.9442 ± 0.0099 | 0.7670 ± 0.0343 |
| 17 | ResNet50V2 | 0.7548 ± 0.0283 | 0.9081 ± 0.0106 | 0.7824 ± 0.0395 | 0.9358 ± 0.0112 | 0.7379 ± 0.0346 |
| 18 | VGG16 | 0.7414 ± 0.0226 | 0.8986 ± 0.0100 | 0.7984 ± 0.0288 | 0.9208 ± 0.0102 | 0.6957 ± 0.0280 |
| 19 | VGG19 | 0.6639 ± 0.0309 | 0.8749 ± 0.0120 | 0.6499 ± 0.0731 | 0.9248 ± 0.0136 | 0.6623 ± 0.0341 |
| 20 | Xception | 0.7806 ± 0.0291 | 0.9179 ± 0.0113 | 0.8013 ± 0.0354 | 0.9436 ± 0.0109 | 0.7684 ± 0.0377 |

## 4.3. Classification of Other Pulmonary Infections

Table 5 summarizes the performance for all trained models in classifying images of other pulmonary infections. EfficientNetB5 attained the highest F1 score and accuracy, DenseNet121 obtained the greatest sensitivity, and EfficientNetB6 got the largest specificity and precision.

Table 5: Summary of model performance for classification of non-Covid pulmonary infections. The best performing model in each metric is highlighted in green. EfficientNetB5 attained the highest F1 score and accuracy, DenseNet201 the greatest sensitivity, and EfficientNetB6 the highest specificity and precision.

| # | Model | F1 | Accuracy | Sensitivity | Specificity | Precision |
|---|---|---|---|---|---|---|
| 1 | DenseNet121 | 0.8239 ± 0.0242 | 0.8966 ± 0.0131 | 0.8188 ± 0.0313 | 0.9293 ± 0.0100 | 0.8315 ± 0.0234 |
| 2 | DenseNet169 | 0.8245 ± 0.0244 | 0.8971 ± 0.0128 | 0.8188 ± 0.0313 | 0.9301 ± 0.0108 | 0.8333 ± 0.0243 |
| 3 | DenseNet201 | 0.8262 ± 0.0256 | 0.8989 ± 0.0133 | 0.8178 ± 0.0344 | 0.9325 ± 0.0091 | 0.8374 ± 0.0213 |
| 4 | EfficientNetB0 | 0.8207 ± 0.0249 | 0.8980 ± 0.0129 | 0.7976 ± 0.0300 | 0.9401 ± 0.0088 | 0.8483 ± 0.0233 |
| 5 | EfficientNetB1 | 0.7482 ± 0.0535 | 0.8623 ± 0.0244 | 0.7210 ± 0.0571 | 0.9221 ± 0.0180 | 0.7889 ± 0.0463 |
| 6 | EfficientNetB2 | 0.8121 ± 0.0246 | 0.8931 ± 0.0127 | 0.7901 ± 0.0311 | 0.9363 ± 0.0095 | 0.8398 ± 0.0233 |
| 7 | EfficientNetB3 | 0.8170 ± 0.0243 | 0.8952 ± 0.0129 | 0.7979 ± 0.0310 | 0.9360 ± 0.0099 | 0.8414 ± 0.0234 |
| 8 | EfficientNetB4 | 0.8288 ± 0.0247 | 0.9009 ± 0.0138 | 0.8143 ± 0.0301 | 0.9373 ± 0.0131 | 0.8496 ± 0.0280 |
| 9 | EfficientNetB5 | 0.8385 ± 0.0278 | 0.9077 ± 0.0140 | 0.8172 ± 0.0367 | 0.9458 ± 0.0084 | 0.8643 ± 0.0225 |
| 10 | EfficientNetB6 | 0.8157 ± 0.0200 | 0.8963 ± 0.0103 | 0.7747 ± 0.0273 | 0.9483 ± 0.0064 | 0.8648 ± 0.0166 |
| 11 | EfficientNetB7 | 0.8038 ± 0.0210 | 0.8856 ± 0.0111 | 0.7905 ± 0.0277 | 0.9262 ± 0.0106 | 0.8235 ± 0.0235 |
| 12 | InceptionResNetV2 | 0.7919 ± 0.0239 | 0.8790 ± 0.0123 | 0.7798 ± 0.0303 | 0.9210 ± 0.0099 | 0.8073 ± 0.0245 |
| 13 | InceptionV3 | 0.7963 ± 0.0286 | 0.8824 ± 0.0150 | 0.7799 ± 0.0367 | 0.9254 ± 0.0123 | 0.8177 ± 0.0275 |
| 14 | ResNet101V2 | 0.7837 ± 0.0279 | 0.8717 ± 0.0163 | 0.7818 ± 0.0311 | 0.9101 ± 0.0163 | 0.7900 ± 0.0340 |
| 15 | ResNet152V2 | 0.7835 ± 0.0254 | 0.8766 ± 0.0123 | 0.7600 ± 0.0381 | 0.9258 ± 0.0115 | 0.8154 ± 0.0244 |
| 16 | ResNet50 | 0.8177 ± 0.0241 | 0.8933 ± 0.0128 | 0.8114 ± 0.0323 | 0.9279 ± 0.0096 | 0.8272 ± 0.0232 |
| 17 | ResNet50V2 | 0.7697 ± 0.0275 | 0.8668 ± 0.0138 | 0.7552 ± 0.0355 | 0.9137 ± 0.0118 | 0.7888 ± 0.0273 |
| 18 | VGG16 | 0.6865 ± 0.0299 | 0.8240 ± 0.0136 | 0.6544 ± 0.0392 | 0.8959 ± 0.0146 | 0.7304 ± 0.0305 |
| 19 | VGG19 | 0.6346 ± 0.0303 | 0.7840 ± 0.0162 | 0.6011 ± 0.0692 | 0.8613 ± 0.0208 | 0.6535 ± 0.0284 |
| 20 | Xception | 0.8017 ± 0.0299 | 0.8854 ± 0.0159 | 0.7863 ± 0.0360 | 0.9268 ± 0.0122 | 0.8210 ± 0.0289 |

## 4.4. Testing on Alternative Dataset

To determine the capacity of the model to scale to different datasets, the top performing model (EfficientNetB5 for most tasks) was additionally deployed on a secondary dataset from Kaggle: COVID-CTset (Rahimzadeh Mohammad et al., 2020), a large dataset containing 63849 CT images from 377 patients (96 COVID-19 positive and 283 Covid-19 negative). To facilitate faster testing, we considered a subset of the data comprised of 12058 images from those 377 patients. On this newer dataset the model obtained an accuracy of 0.9845 ± 0.0109, F1 score of 0.9599 ± 0.0251, sensitivity of 0.9682 ± 0.0099, specificity of 0.9883 ± 0.0150, and precision of 0.9526 ± 0.0523. Note, the higher performance on the alternate dataset, although shocking, is expected as it involves a binary classification task, which is naturally a lot easier than a multi-class classification task. This is particularly true for COVID-19, where distinguishing between a lung infected with COVID-19 related pneumonia and one with a community acquired pneumonia might be difficult. This also demonstrates the scalability of our proposed network, suggesting it may adapt to a wide variety of image types.

### 4.5. Brief Note on Algorithm Execution Time

2 tests were conducted to examine the execution time of the proposed model (EfficientNetB5) for classification. The first test was conducted on a Tesla V100-SXM2-32GB GPU with 5120 cores and 32 GB of RAM. This was the system used for model training. The second was done on a personal household computer: Dell XPS 13 with 16 GB of RAM and Intel(R) Core(TM) i7-10710U CPU @ 1.10GHz processor. For each device, the model was used to predict 1000 images. The Tesla V100-SXM2-32GB GPU took a total of 80.897427 seconds for prediction with an average of 0.0808974 seconds/image. The Dell XPS 13 computer took a total of 490.48266 seconds for prediction with an average of 0.49048266 seconds/image. The model scripts were configured to be compatible with both GPU and CPU based environments with simply a one word change in keyword arguments, providing seamless integrability regardless of computational resources available.

### 4.6. Brief Note on Algorithm Training Time

All models were trained on a Tesla V100-SXM2-32GB GPU for a total of 25 epochs. Each epoch of training took approximately 140 seconds when training on 3336 images belonging to 3 different classes. Hence depending on the number of epochs chosen, the model may take anywhere from 20 minutes to 3 hours to train. Note: although we replicated our training process for 25 rounds to gain statistical significance and confidence in our predictions, hospitals need not conduct similar analyses. In cases where a hospital does not have the computational resources needed to train the models themselves, they may easily outsource training to computational clusters like Google Cloud, Microsoft Azure, or Amazon AWS for a marginal cost. Once trained, the model states and weights can easily be downloaded both in its native format or in a lightweight TensorFlow Lite format, which may be easily embedded on edge devices with limited computational resources. Once embedded, the model does offer high performance speeds as discussed in section 4.5.

## 5. GradCAM Visualization

This section visualizes the GradCAMs, comparing them between COVID-19 positive patients and COVID-19 negative patients. It is evident from figure 1 that the networks indeed focused upon features radiologically recognized as being suggestive of lung involvement in COVID-19 in cases with high pre-test probabilities for making classifications with high accuracy. These images are generated for EfficientNet-B5, DenseNet169, ResNet50, InceptionV3, Xception, and VGG16, for the best performing models in each family. We note that while the generated heatmaps were specific and localized when visualizing COVID-19, they tended to become more diffused and spread out for almost all models when visualizing non-COVID-19 images.

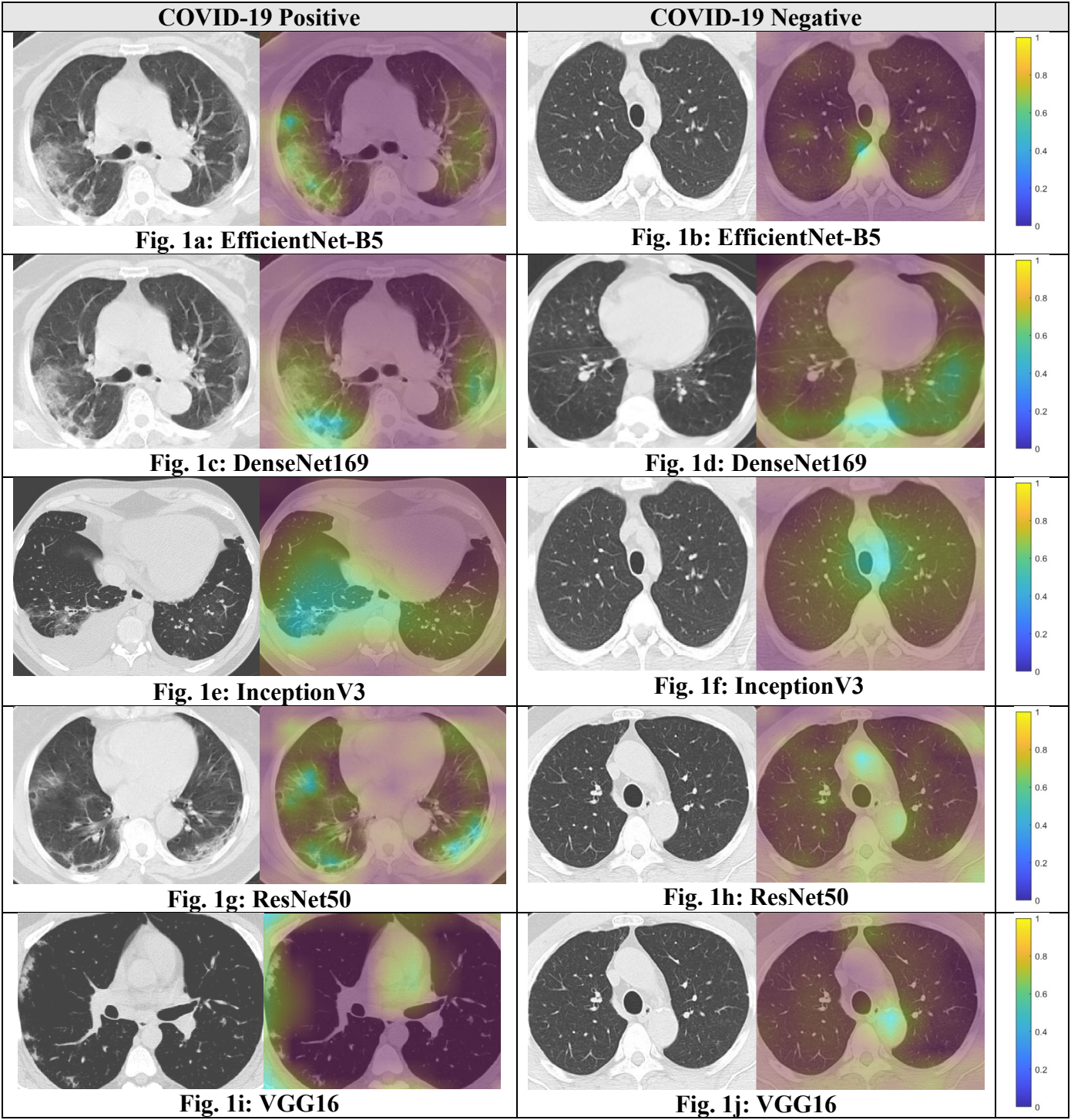

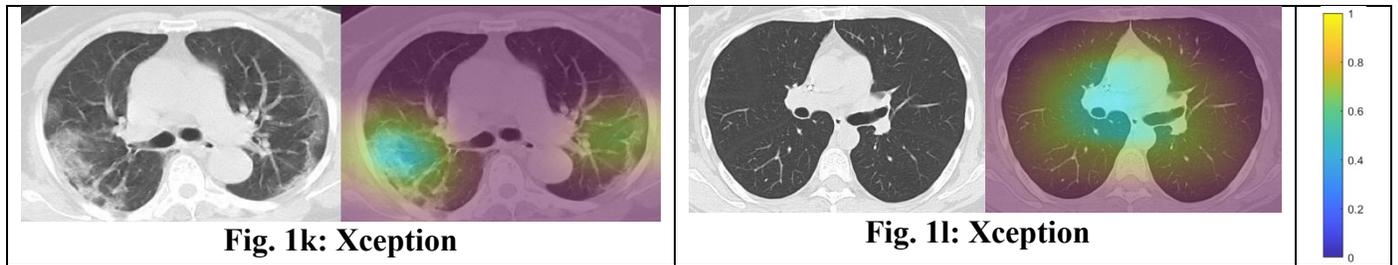

*Figure 1: Comparison of GradCAM Visualizations between COVID-19 Positive and COVID-19 negative images. We see that the saliency maps are much more diffuse and spread out throughout the image in the case of non-COVID images, which suggests that the model was unable to pinpoint regions that would hint towards the presence of COVID-19.*

## 6. Intermediate Activation Maps

Figure 2 shows the complexity of these maps that evolve with time. In the early stages of the network (Fig. 2b, 2c), the model starts to learn basic feature maps from the image, such as their edges, color gradients, etc. In layers near the middle of these models (Fig. 2d, 2e, 2f), these maps get significantly more complicated, picking up on ground-glass opacities, consolidations, and crazy paving patterns as noted in the highlighted sections of greater vibrance. Gradually, these embeddings become more and more complex, veering from being human interpretable towards features only understandable by a computer (Fig 2g, 2h). In the later stages of network (Fig 2i, 2j, 2k), the image complexity appears to stagnate, and the model starts reconstructing images that resemble the input. This functionality hints towards generative activity within the model, where the reconstructed images more dominantly represent characteristics useful in diagnosing COVID-19. Moreover, the clarity in maps even farther into the network reaffirms that our model isn't overfitting strongly, for in such a case, final maps would be entirely black (suggesting no activity) or stop evolving at all. Nonetheless, the lower rates of map evolution post layer 450 does hint towards mild overfitting. Although additional work is required to fully establish the trends between activation maps and overfitting, including optimal policies for stopping training, this study establishes precedent for such work. Overall, these maps with their invaluable insight show how the model learns, playing an equally important role as GradCAMs, the current visualization scheme of choice.

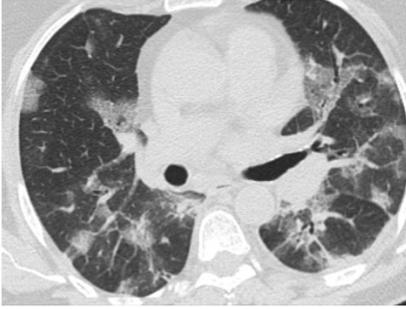

| Input Image<br>Fig. 2a. | Block2a_se_excite<br>Fig. 2b. Layer 48 | Block2e_activation<br>Fig. 2c. Layer 101 |
| --- | --- | --- |
| Block3c_drop<br>Fig. 2d. Layer 152 | Block4b_activation<br>Fig. 2e. Layer 202 | Block4e_se_expand<br>Fig. 2f. Layer 251 |
| Block5c_se_expand<br>Fig. 2g. Layer 324 | Block5g_project_conv<br>Fig. 2h. Layer 386 | Block6d_bn<br>Fig. 2i. Layer 437 |

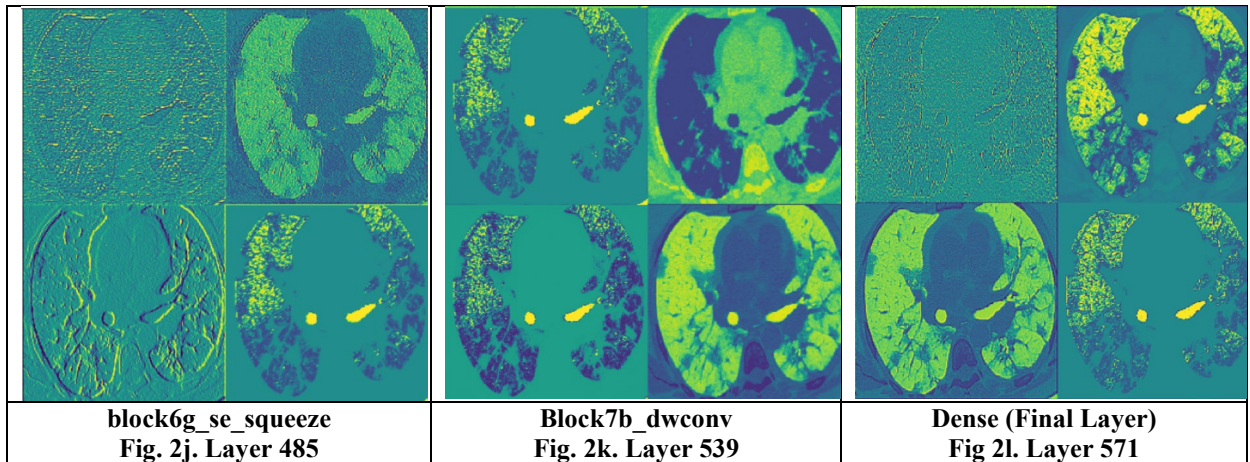

| block6g_se_squeeze | Block7b_dwconv | Dense (Final Layer) |
| Fig. 2j. Layer 485 | Fig. 2k. Layer 539 | Fig 2l. Layer 571 |

*Figure 2: Intermediate Activation Maps: From 4-image sub slices from the intermediate activation maps, we can see the progression of the model's learning behavior. The model progresses from maps similar to input images during its early layers (Fig 2b.) towards maps of increasing complexity as the layer depth increases (Fig 2f, 2g, 2h). Near the final layers of the model, the maps hint towards generative activity as the model constructs images similar to the input image from largely simplified pixelated maps that precede it. Fig 6.l represents the final activation map used before the model prediction. We can note the model ability to close-in upon and depict small pixels and voxels in the input image.*

Next, we compare the activation maps between COVID-19 positive and negative images. We note that the activation maps for COVID-positive images tend to focus on more intricate lung patterns like paving patterns, ground class opacities, and consolidations, whereas the feature maps for COVID-19 negative cases are less detailed and have a more uniform pattern. The maps become increasingly localized at later layers, indicating that the model closely examines each minute pattern and opacity on the lung before classifying a patient.

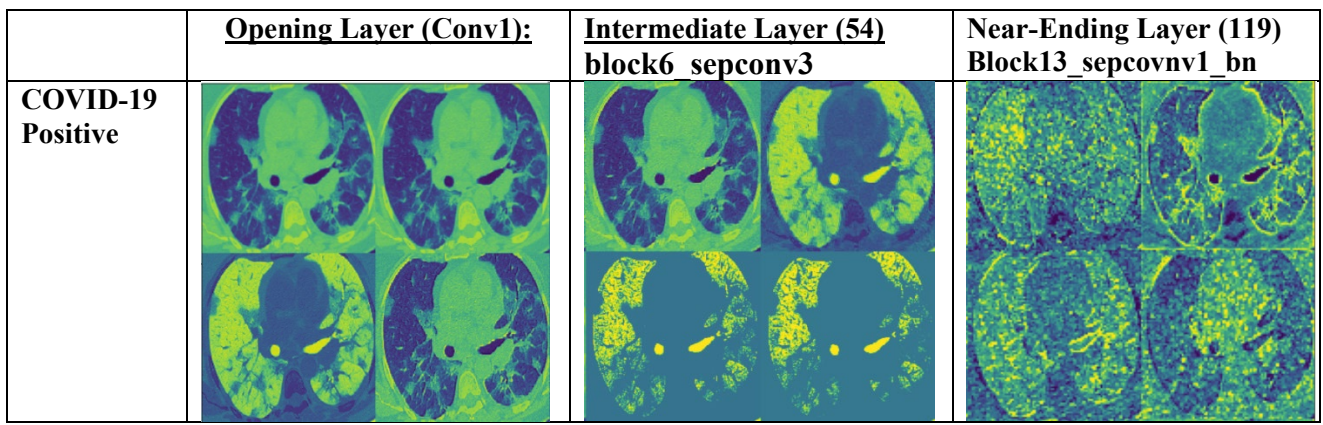

|  | **Opening Layer (Conv1):** | **Intermediate Layer (54)** block6_sepconv3 | **Near-Ending Layer (119)** Block13_sepcovnv1_bn |
| **COVID-19 Positive** | | | |

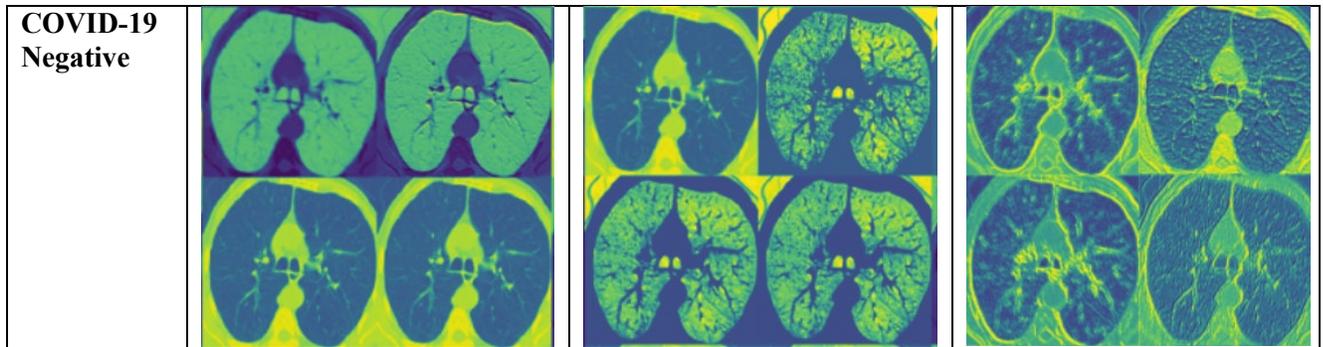

*Figure 3: Comparison of intermediate activate maps for COVID-19 positive versus COVID-19 negative images. We note that while the activation patterns for COVID-19 positive images tend to capture lung opacities, the patterns for Non-COVID images are a bit simpler and uniform. Once again, near the later layers of the model, the feature maps become increasingly uninterpretable.*

## 7. PROPOSED NETWORK ARCHITECTURE

The entire EfficientNet-B5 architecture is 571 layers long and is summarized through the use of modules and sub-blocks in figure 4. At the core of EfficientNet-B5 rest the following characteristics: (1) a highly accurate and efficient architecture found by performing a neural architecture search via the AutoML MNAS framework (Tan et al., 2018). (2) A strategic and low latency use of the mobile inverted bottleneck convolution similar to that in MobileNet (Howard et al., 2017). (3) A systematic compound model scaling approach for maximizing performance gains within strict resource and computational limitations (Tan & Le, 2019). Altogether, these features allow ML practitioners to optimize small networks using neural architecture search, which saves time and resources, and then scale them accordingly to provide higher performance. These characteristics allow for both a computational efficient network architecture and high accuracy, qualities that are both essential given the low computation resources available in hospitals and need for high model performance to reduce false positive/negatives.

Furthermore, the enhancements in the proposed architecture stem from a systematic use of neural architecture search and compound scaling, qualities which may be applied for any field like brain tumor diagnosis and facial recognition, not just COVID-19 diagnosis. This is reflected in our performance results on alternate datasets as well as in EfficientNetsB5's extraordinary performance on the ImageNet datasets. More information regarding EfficientNets and equations used for upscaling can be found in the Appendix.

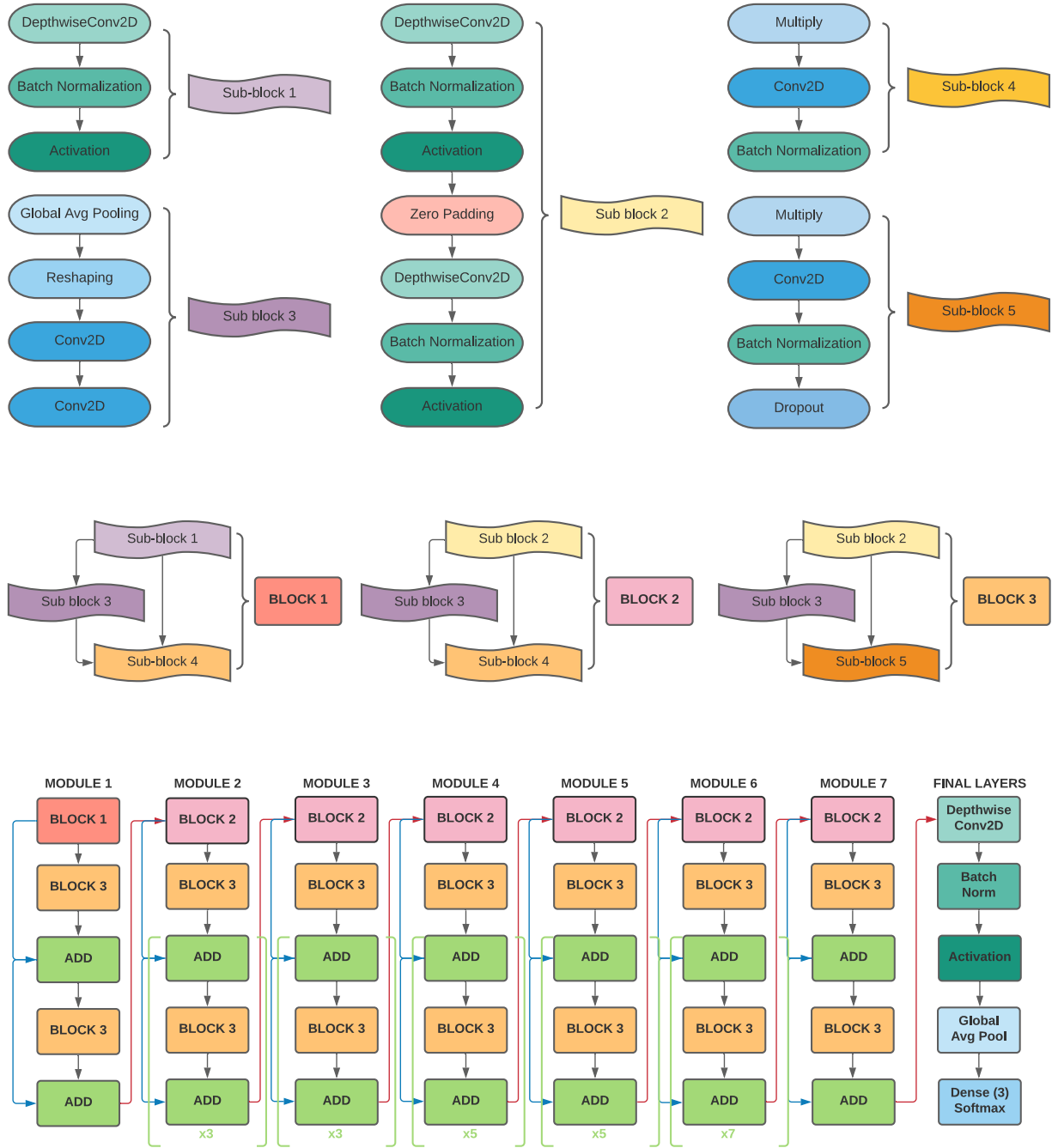

*Figure 4: Summary of EfficientNetB5 Architecture*

## 8. DISCUSSION

The CNNs showed a large increase in the sensitivity of COVID-19 diagnosis using CT scans (0.9769 for a 3-class classification test) in comparison to RT-PCR (0.71 for a simpler binary-classification test) (Fang et al., 2020). As proposed, the Efficient-Net family performed exceedingly well for diagnosis, attaining 6 out of the top 9 F1 scores amongst all models for COVID-19 classification, 7 out of the top 8 F1 scores for classifying healthy images, and 7 out of the top 11 F1 scores for classifying images of other pulmonary infections.

The proposed model of choice, EfficientNet-B5 obtained the highest F1 score, accuracy, and sensitivity for COVID-19 positive images; the highest F1 score, accuracy, sensitivity, specificity, and precision for healthy images, and the highest F1 score and accuracy for images of other pulmonary infections. It was consistently the most balanced classifier as shown by its consistently highest F1 scores and showed consistently high levels of performance for all classification categories. In an environment dominated with COVID-19, mitigating secondary infections is critical and one must ensure an infected patient isn't categorized as healthy, hence sensitivity was considered an extremely important metric. With a sensitivity of $0.9788 \pm 0.0055$, EfficientNet-B5's performance was significantly higher than all other models considered within a confidence level of 95%. The high-performance gains from EfficientNetB5 and its computational effectiveness strongly advance our proposal for a more uniform usage of EfficientNets for COVID-19 classification tasks.

Other notable models from the EfficientNet class are EfficientNetB0 and EfficientNetB5. The former, EfficientNetB0, although the smallest model in size (43.24 MB) obtained the 2[nd] highest F1 scores for COVID-19 classification amongst the EfficientNet family and 5[th] highest overall; 5[th] highest F1 score for healthy image classification overall; and 6[th] highest F1 score overall for classifying other pulmonary infections. Given that these machine learning models are to be deployed in hospitals with potentially limited computational resources, EfficientNetB0 thus provides a network simple to deploy on edge devices with limited computing resources.

Although EfficientNet-B5 was consistently the most balanced classifier, we would be remiss to not acknowledge the potential of DenseNets, which offered two significant advantages: high model performance with low network size. DenseNet121, DenseNet169, and DenseNet201 had sizes of 81.8 MB, 146.3 MB, and 211.21 MB, respectively, significantly lower the current predominantly used model of choice: ResNet50 (270.54 MB). Moreover, they outperformed ResNet50 in every single classification metric (F1 score, accuracy, sensitivity, specificity, and precision) and category (COVID-19 positive, healthy, other pulmonary infections). Finally, DenseNet201 was model with the highest sensitivity ($0.8188 \pm 0.0313$) in classifying other pulmonary infections and DenseNet169 had the highest specificity ($0.9759 \pm 0.0072$) and precision ($0.9779 \pm 0.0063$) for COVID-19 positive images. Altogether, DenseNets are another promising yet underutilized class for COVID-19 diagnosis.

Our use of GradCAMs and Intermediate Activation Maps, we increased the interpretability of our model. In particular, our use of GradCAMs allowed us to hone in on and understand which portions of input images were most essential to model classification. On the other hand, intermediate activation maps allowed users to "peek" into CNN's black-box hood and understand intermediary steps that it was taking, an invaluable tool for both debugging networks and destigmatizing their use. Finally, the performance of our model on an alternative dataset and its computational time (~0.1s on GPUs and ~0.5s on CPUs) were both positive indicators of its applications. Particularly, its ability to perform on different dataset formats and provide rapid, accurate results speaks to its generalizability and larger avenues for deployment.

Although the models developed in this paper can offer high accuracies for COVID-19 diagnosis, several technological, economic, medical, and psychological factors must be considered before deploying them in real-time environments. First, although our network modifications and visualization schemes allow for greater insight into model learning, deep learning still holds elements its black-box nature. Thus, it is hard to pinpoint changes in model performance from changes in network structure and weights. This can make generalization and scalability to machines universally challenging. Secondly, CT scans cost significantly higher than a regular RT-PCR test. This consideration may make it cost-prohibitive and difficult to include in everyday practices. Thirdly, CT has an unspoken radiation cost associated with it. Exposure to Chest CT greatly increases one's chances of acquiring cancer in subsequent years and therefore is not always an optimal choice of testing. Fourthly, deployment of Artificial Intelligence based systems in hospitals is routinely critiqued and stigmatized because of their potential psychological impacts on individuals. Doctors not only help diagnose patients, but also offer them emotional support and pragmatic advice on dealing with the disease. The currently designed machine learning models, however, only offer a numerical output, which can be demoralizing for patients without proper guidance from doctors and clinicians. Finally, some families may be skeptical of results obtained from a computer-based system versus a doctor with years of experience. As a result, while the current model is a useful tool in assisting doctors in diagnosing, it currently only serves as a proof of concept and as an enhancement to current approaches.

Given these factors, although the results of this study are useful in an academic setting, a close examination of a location's health system may be required before deploying them in hospitals. Particularly for countries with sufficient PCR-testing kits like the US, diagnosis should be deferred to RT-PCR, while ML-based CT imaging can play useful roles in assessing the severity of patients' cases. However, in countries with severe PCR testing deficiencies, CT has already become a mode of diagnosis, as the risks of letting COVID-19 spread have been deemed more pertinent than the risks of potential cancer risks several years down the line. For such patients, our proposed system not only helps alleviate potential financial burdens on individuals from having to sponsor radiologist fees, but also offers radiologists a useful tool for reducing their workloads.

## 9. CONCLUSION

In summary, our project presents a thorough analysis of This study successfully presented a thorough analysis of the use of traditional and custom ML techniques, specifically CNN architectures, for COVID-19 detection based on chest CT. Our results highlighted the potential for using EfficientNets and DenseNets for COVID-19 diagnosis purposes, encouraging their use in an environment where they are vastly underutilized. With the backlog of PCR tests, causing results to take anywhere from 6 hours to 4 days to reach patients, our work presents an optimized machine learning framework for streamlining that bottleneck, helping control the spread of the disease and bringing the world towards normalcy.

## 10. FUNDING

This work was supported by the National Science Foundation under Award Number 2027456 (COVID-ARC).

## 11. REFERENCES

Alrahhal, M., & K P, S. (2021). COVID-19 Diagnostic System Using Medical Image Classification and Retrieval: A Novel Method for Image Analysis. *The Computer Journal*, *00*(00), 2021. https://doi.org/10.1093/COMJNL/BXAB051

Ardakani, A. A., Kanafi, A. R., Acharya, U. R., Khadem, N., & Mohammadi, A. (2020). Application of deep learning technique to manage COVID-19 in routine clinical practice using CT images: Results of 10 convolutional neural networks. *Computers in Biology and Medicine*, *121*, 103795. https://doi.org/10.1016/j.compbiomed.2020.103795

Bougourzi, F., Contino, R., Distante, C., & Taleb-Ahmed, A. (2021). *CNR-IEMN: A Deep Learning Based Approach to Recognise Covid-19 from CT-Scan*. 8568–8572. https://doi.org/10.1109/ICASSP39728.2021.9414185

Butt, C., Gill, J., Chun, D., & Babu, B. A. (2020). Deep learning system to screen coronavirus disease 2019 pneumonia. *Applied Intelligence*. https://doi.org/10.1007/s10489-020-01714-3

Chaudhary, P. K., & Pachori, R. B. (2021). FBSED based automatic diagnosis of COVID-19 using X-ray and CT images. *Computers in Biology and Medicine*, *134*, 104454. https://doi.org/10.1016/J.COMPBIOMED.2021.104454

Chollet, Francois. (2017). Visualizing convnet filters. In *Deep Learning with Python* (1st ed., pp. 160–172). Manning Publications Co.

Chollet, François. (2017). Xception: Deep learning with depthwise separable convolutions. *Proceedings - 30th IEEE Conference on Computer Vision and Pattern Recognition, CVPR 2017*, *2017-January*, 1800–1807. https://doi.org/10.1109/CVPR.2017.195

Chowdhury, N. K., Kabir, M. A., Rahman, M. M., & Rezoana, N. (2021). ECOVNet: a highly effective ensemble based deep learning model for detecting COVID-19. *PeerJ Computer Science*, *7*, e551. https://doi.org/10.7717/PEERJ-CS.551

Deng, J., Dong, W., Socher, R., Li, L.-J., Kai Li, & Li Fei-Fei. (2010). *ImageNet: A large-scale hierarchical image database*. 248–255. https://doi.org/10.1109/CVPR.2009.5206848

Dong, D., Tang, Z., Wang, S., Hui, H., Gong, L., Lu, Y., Xue, Z., Liao, H., Chen, F., Yang, F., Jin, R., Wang, K., Liu, Z., Wei, J., Mu, W., Zhang, H., Jiang, J., Tian, J., & Li, H. (2020). The role of imaging in the detection and management of COVID-19: a review. *IEEE Reviews in Biomedical Engineering*, *PP*. https://doi.org/10.1109/RBME.2020.2990959

Fang, Y., Zhang, H., Xie, J., Lin, M., Ying, L., Pang, P., & Ji, W. (2020). Sensitivity of chest CT for COVID-19: Comparison to RT-PCR. In *Radiology* (Vol. 296, Issue 2, pp. E115–E117). Radiological Society of North America Inc. https://doi.org/10.1148/radiol.2020200432

Foysal, M., & Aowlad Hossain, A. B. M. (2021). *COVID-19 Detection from Chest CT Images using Ensemble Deep Convolutional Neural Network*. 1–6. https://doi.org/10.1109/INCET51464.2021.9456387

Garain, A., Basu, A., Giampaolo, F., Velasquez, J. D., & Sarkar, R. (2021). Detection of COVID-19 from CT scan images: A spiking neural network-based approach. *Neural Computing and Applications 2021*, 1–14. https://doi.org/10.1007/S00521-021-05910-1

Garg, P., Ranjan, R., Upadhyay, K., Agrawal, M., & Deepak, D. (2021). *Multi-Scale Residual Network for Covid-19 Diagnosis Using Ct-Scans*. 8558–8562. https://doi.org/10.1109/ICASSP39728.2021.9414426

He, K., Zhang, X., Ren, S., & Sun, J. (2016a). Identity Mappings in Deep Residual Networks. *Lecture Notes in Computer Science (Including Subseries Lecture Notes in Artificial Intelligence and Lecture Notes in Bioinformatics)*, *9908 LNCS*, 630–645. https://doi.org/10.1007/978-3-319-46493-0_38

He, K., Zhang, X., Ren, S., & Sun, J. (2016b). Deep residual learning for image recognition. *Proceedings of the IEEE Computer Society Conference on Computer Vision and Pattern Recognition*, *2016-December*, 770–778. https://doi.org/10.1109/CVPR.2016.90

Howard, A. G., Zhu, M., Chen, B., Kalenichenko, D., Wang, W., Weyand, T., Andreetto, M., & Adam, H. (2017). *MobileNets: Efficient Convolutional Neural Networks for Mobile Vision Applications*.

Huang, G., Liu, Z., van der Maaten, L., & Weinberger, K. Q. (2016). Densely Connected Convolutional Networks. *Proceedings - 30th IEEE Conference on Computer Vision and Pattern Recognition, CVPR 2017*, *2017-January*, 2261–2269.

Ibrahim, M. R., Youssef, S. M., & Fathalla, K. M. (2021). Abnormality detection and intelligent severity assessment of human chest computed tomography scans using deep learning: a case study on SARS-COV-2 assessment. *Journal of Ambient Intelligence and Humanized Computing 2021*, *1*, 1–24. https://doi.org/10.1007/S12652-021-03282-X

Islam, M. M., Karray, F., Alhajj, R., & Zeng, J. (2021). A Review on Deep Learning Techniques for the Diagnosis of Novel Coronavirus (COVID-19). *IEEE Access*, *9*, 30551–30572. https://doi.org/10.1109/ACCESS.2021.3058537

Jin, C., Chen, W., Cao, Y., Xu, Z., Tan, Z., Zhang, X., Deng, L., Zheng, C., Zhou, J., Shi, H., & Feng, J. (2020). Development and evaluation of an artificial intelligence system for COVID-19 diagnosis. *Nature Communications*, *11*(1), 5088. https://doi.org/10.1038/s41467-020-18685-1

Kamel, M. A., Abdelshafy, M., Abdulrazek, M., Abouelkhir, O., Fawzy, A., & Sahlol, A. T. (2021). Efficient Classification Approach Based on COVID-19 CT Images Analysis with Deep Features. *Proceedings - IEEE 2021 International Conference on Computing, Communication, and Intelligent Systems, ICCCIS 2021*, 459–464. https://doi.org/10.1109/ICCCIS51004.2021.9397189

Karen Simonyan∗ & Andrew Zisserman+. (2018). VERY DEEP CONVOLUTIONAL NETWORKS FOR LARGE-SCALE IMAGE RECOGNITION Karen. *American Journal of Health-System Pharmacy*, *75*(6), 398–406.

Kaya, A., Atas, K., & Myderrizi, I. (2021). Implementation of CNN based COVID-19 classification model from CT images. *SAMI 2021 - IEEE 19th World Symposium on Applied Machine Intelligence and Informatics, Proceedings*, 201–206. https://doi.org/10.1109/SAMI50585.2021.9378646

Kim, H., Hong, H., & Ho Yoon, S. (2020). Diagnostic performance of ct and reverse transcriptase polymerase chain reaction for coronavirus disease 2019: A meta-analysis. *Radiology*, *296*(3), E145–E155. https://doi.org/10.1148/radiol.2020201343

Lalmuanawma, S., Hussain, J., & Chhakchhuak, L. (2020). Applications of machine learning and artificial intelligence for Covid-19 (SARS-CoV-2) pandemic: A review. *Chaos, Solitons and Fractals*, *139*, 110059. https://doi.org/10.1016/j.chaos.2020.110059

Li, X., Tan, W., Liu, P., Zhou, Q., & Yang, J. (2021). Classification of COVID-19 Chest CT Images Based on Ensemble Deep Learning. *Journal of Healthcare Engineering*, *2021*. https://doi.org/10.1155/2021/5528441

Marques, G., Agarwal, D., & de la Torre Díez, I. (2020). Automated medical diagnosis of COVID-19 through EfficientNet convolutional neural network. *Applied Soft Computing Journal*, *96*. https://doi.org/10.1016/j.asoc.2020.106691

Muftuoglu, Z., Kizrak, M. A., & Yildlnm, T. (2020, August 1). Differential Privacy Practice on Diagnosis of COVID-19 Radiology Imaging Using EfficientNet. *INISTA 2020 - 2020 International Conference on INnovations in Intelligent SysTems and Applications, Proceedings*. https://doi.org/10.1109/INISTA49547.2020.9194651

Oyelade, O. N., Ezugwu, A. E., & Chiroma, H. (2021). CovFrameNet: An enhanced deep learning framework for COVID-19 detection. *IEEE Access*. https://doi.org/10.1109/ACCESS.2021.3083516

Ozturk, T., Talo, M., Yildirim, E. A., Baloglu, U. B., Yildirim, O., & Rajendra Acharya, U. (2020). Automated detection of COVID-19 cases using deep neural networks with X-ray images. *Computers in Biology and Medicine*, *121*. https://doi.org/10.1016/j.compbiomed.2020.103792

[dataset] Rahimzadeh Mohammad, Sakhaei Seyed Mohammad, & Attar Abolfazl. (2020). *COVID-CTset : A Large COVID-19 CT Scans dataset containing 63849 images from 377 patients*. https://github.com/mr7495/COVID-CTset

Selvaraju, R. R., Cogswell, · Michael, Das, A., Ramakrishna Vedantam, ·, Parikh, · Devi, Dhruv Batra, ·, Cogswell, M., Vedantam, R., & Parikh, D. (2020). Grad-CAM: Visual Explanations from Deep Networks via Gradient-Based Localization. *International Journal of Computer Vision*, *128*, 336–359. https://doi.org/10.1007/s11263-019-01228-7

Singh, V. K., & Kolekar, M. H. (2021). Deep learning empowered COVID-19 diagnosis using chest CT scan images for collaborative edge-cloud computing platform. *Multimedia Tools and Applications 2021*, 1–28. https://doi.org/10.1007/S11042-021-11158-7

[dataset] Soares, Eduardo (Universidad de Lancaster); Angelov, P. (Universidad de L. (2020). *A COVID multiclass dataset of CT scans*. https://doi.org/10.34740/KAGGLE/DSV/1235046

Szegedy, C., Ioffe, S., Vanhoucke, V., & Alemi, A. A. (2017). Inception-v4, inception-ResNet and the impact of residual connections on learning. *31st AAAI Conference on Artificial Intelligence, AAAI 2017*, 4278–4284.

Szegedy, C., Liu, W., Jia, Y., Sermanet, P., Reed, S., Anguelov, D., Erhan, D., Vanhoucke, V., & Rabinovich, A. (2015). Going deeper with convolutions. *Proceedings of the IEEE Computer Society Conference on Computer Vision and Pattern Recognition*, *07-12-June-2015*, 1–9. https://doi.org/10.1109/CVPR.2015.7298594

Szegedy, C., Vanhoucke, V., Ioffe, S., Shlens, J., & Wojna, Z. (2016). Rethinking the Inception Architecture for Computer Vision. *Proceedings of the IEEE Computer Society Conference on Computer Vision and Pattern Recognition*, *2016-December*, 2818–2826. https://doi.org/10.1109/CVPR.2016.308

Tan, M., Chen, B., Pang, R., Vasudevan, V., Sandler, M., Howard, A., & Le, Q. V. (2018). MnasNet: Platform-Aware Neural Architecture Search for Mobile. *Proceedings of the IEEE Computer Society Conference on Computer Vision and Pattern Recognition*, *2019-June*, 2815–2823.

Tan, M., & Le, Q. V. (2019). EfficientNet: Rethinking Model Scaling for Convolutional Neural Networks. *36th International Conference on Machine Learning, ICML 2019*, *2019-June*, 10691–10700.

Waleed Salehi, A., Baglat, P., & Gupta, G. (2020). Review on Machine and Deep Learning Models for the Detection and Prediction of Coronavirus. *Materials Today: Proceedings*. https://doi.org/10.1016/j.matpr.2020.06.245

Wang, B., Jin, S., Yan, Q., Xu, H., Luo, C., Wei, L., Zhao, W., Hou, X., Ma, W., Xu, Z., Zheng, Z., Sun, W., Lan, L., Zhang, W., Mu, X., Shi, C., Wang, Z., Lee, J., Jin, Z., … Dong, J. (2021). AI-assisted CT imaging analysis for COVID-19 screening: Building and deploying a medical AI system. *Applied Soft Computing*, *98*, 106897. https://doi.org/10.1016/J.ASOC.2020.106897

Wu, X., Hui, H., Niu, M., Li, L., Wang, L., He, B., Yang, X., Li, L., Li, H., Tian, J., & Zha, Y. (2020). Deep

learning-based multi-view fusion model for screening 2019 novel coronavirus pneumonia: A multicentre study. *European Journal of Radiology*, *128*, 109041. https://doi.org/10.1016/j.ejrad.2020.109041

Wu, Y. H., Gao, S. H., Mei, J., Xu, J., Fan, D. P., Zhang, R. G., & Cheng, M. M. (2021). JCS: An Explainable COVID-19 Diagnosis System by Joint Classification and Segmentation. *IEEE Transactions on Image Processing*, *30*, 3113–3126. https://doi.org/10.1109/TIP.2021.3058783

Xiong, Z., Wang, R., Bai, H. X., Halsey, K., Mei, J., Li, Y. H., Atalay, M. K., Jiang, X. L., Fu, F. X., Thi, L. T., Huang, R. Y., Liao, W. H., Pan, I., Choi, J. W., Zeng, Q. H., Hsieh, B., CuiWang, D., Sebro, R., Hu, P. F., … Qi, Z. Y. (2020). Artificial Intelligence Augmentation of Radiologist Performance in Distinguishing COVID-19 from Pneumonia of Other Origin at Chest CT. *Radiology*, *296*(3), E156–E165. https://doi.org/10.1148/radiol.2020201491

Ye, Z., Zhang, Y., Wang, Y., Huang, Z., & Song, B. (2020). Chest CT manifestations of new coronavirus disease 2019 (COVID-19): a pictorial review. *European Radiology*, *30*(8), 4381–4389. https://doi.org/10.1007/s00330-020-06801-0

Yousefzadeh, M., Esfahanian, P., Movahed, S. M. S., Gorgin, S., Lashgari, R., Rahmati, D., Kiani, A., Kahkouee, S., Nadji, S. A., Haseli, S., Hoseinyazdi, M., Roshandel, J., Bandegani, N., Danesh, A., Bakhshayesh Karam, M., & Abedini, A. (2020). ai-corona: Radiologist-Assistant Deep Learning Framework for COVID-19 Diagnosis in Chest CT Scans. https://doi.org/10.1101/2020.05.04.20082081

## 12. APPENDIX

This section provides a basic overview of the machine learning model families we considered in our paper.

### 12.1. EfficientNets

Proposed by Mingxing Tan and Quoc V. Le (Tan & Le, 2019), EfficientNets have quickly revolutionized the current standing of computer vision, providing not only high accuracy results, but attaining them with computational complexities orders of magnitude (8.4x smaller and 6.1 x faster) lower than the best ConvNets. They proposed a novel ConvNet scaling framework, adjusting the width (the number of channels in network layers), depth (number of layers in the CNN), and resolution (the input image size into the model) systematically. Particularly, if

$$depth, d = \alpha^\phi, width\ w = \beta^\phi; and\ resolution\ r = \gamma^\phi, s.t. \alpha * \beta^2 * \gamma^2 \approx 2 , \quad (8)$$

then the model may be scaled by adjusting based on the available computational resources. The base network for EfficientNet is determined by performing neural architecture search over the

AutoML MNAS Framework. This allows EfficientNets to take advantage of a optimized small network, which is both cost and time efficient to compute and scaling procedures which boost performance. Given the need for rapid diagnosis and limited computational capacities for obtaining them, we hypothesized EfficientNets would be the most promising form of diagnosis.

## 12.2. ResNets

Proposed by Kaiming He et al. (He et al., 2016b), ResNets are arguably the most popular CNN architectures for image recognition tasks today. Their strength, the residual learning framework capable of transmitting gradients despite great depths by skip connections and batch-normalization, has rapidly been applied into numerous modern architectures today. This paper considers four types of ResNet architectures: ResNet50, ResNet50V2, ResNet101V2, and ResNet152V2 (Note: for this paper, InceptionResNetV2 is grouped in the InceptionNet category along with InceptionV3 and Xception)

## 12.3. DenseNets

Proposed by Huang et al., (Huang et al., 2016) DenseNets introduce the idea of connecting every layer to every other layer in a feed-forward fashion. This, in turn, allows them to avoid the vanishing-gradient problem, reduce training parameters, and improve feature transmission (Huang et al., 2016). Given their clever design, they are yet another popular CNN of choice. This study considers 3 DenseNet models: DenseNet121, DenseNet169, and DenseNet201 in order of increasing parameter size.

## 12.4. InceptionNet & Xception

Proposed by Svegedy et al. (Szegedy et al., 2015), InceptionNets advance the concept of building CNNs using blocks instead of just convolutional layers, a framework most modern networks utilize. Moreover, they decomposed convolutional operations into spatially separable ones for improved computational resources utilization, increasing both the depth and width of the model while keeping computational costs static. In a subsequent study by Svegedy et al. (Szegedy et al., 2016), they jointly capitalized on InceptionNet's module-based architecture and the residual connections from ResNets to propose InceptionResNetV2, a powerful model combining the best features from ResNet and InceptionNet. Finally, François Chollet (François Chollet, 2017) expanded upon InceptionNets in his work Xception: Deep Learning with Depthwise Separable Convolutions, replacing inception modules with depthwise separable convolutions (a depthwise convolution and then a pointwise convolution). This work examines two models from the InceptionNet class (InceptionNetV2, InceptionResNetV2) and Xception.

## 12.5. VGG

Proposed by Karen Simonyan and Andrew Zisserman (Karen Simonyan∗ & Andrew Zisserman+, 2018), VGG's primary contribution was to experiment with increasing model depth and seeing its impact on model performance. We trained two forms of VGGs: VGG16 and VGG19.